**Probing intrinsic magnetic phases in low-dimensional nearly twin-free $NiPS_3$ single crystals**

*Yeochan An[#], Heejun Yang[#], Sung Jin Park[#], Giung Park[#], Woonghee Cho, Pyeongjae Park, Seokhwan Yun, Yoshimitsu Kohama, and Je-Geun Park**

# authors with equal contributions
* corresponding author

Y. An, H. Yang, S. J. Park, G. Park, W. Cho, P. Park, J.-G. Park
Center for Quantum Materials, Seoul National University, Seoul 08826, Republic of Korea
Department of Physics and Astronomy, Seoul National University, Seoul 08826, Republic of Korea
E-mail: jgpark10@snu.ac.kr

S. Yun
Korea Atomic Energy Research Institute, Daejeon 34057, Republic of Korea

Y. Kohama
Institute for Solid State Physics, University of Tokyo, Kashiwa, Chiba 277-8581, Japan

Funding: Leading Researcher Program of the National Research Foundation of Korea (Grant no. RS-2020-NR049405), The Core Center Program of the Ministry of Education of the Korean government (2021R1A6C101B418)

Keywords: van der Waals magnets, $NiPS_3$, crystallographic twins, magnetic phase transitions, thermal conductivities, spin-phonon scattering

## Abstract

We report the intrinsic thermal and magnetic properties of the low-dimensional van der Waals (vdW) antiferromagnet $NiPS_3$ and explore its emergent magnetic phases by controlling crystallographic twinning. Using nearly twin-free crystals, we resolve intrinsic properties that are typically obscured by multidomain effects in bulk samples. Magnetization results reveal a highly anisotropic, sharp spin-flop transition, confirming the high domain purity of our crystals. Furthermore, high-precision thermodynamic and transport data reveal a broad fluctuation regime around the Néel temperature ($T_N$ = 157.5 K), with a heat capacity anomaly and a concurrent suppression of thermal conductivity. Field-dependent thermal transport shows a small but distinct contribution from spin-lattice coupling, as evidenced by the dip at the spin-flop transition. We develop a theoretical model to explain these properties reported in this paper, with good agreement between experiment and theory. Our work establishes a definitive baseline for bulk properties of $NiPS_3$ and demonstrates the feasibility of resolving intrinsic anisotropies by addressing crystallographic twinning in vdW magnets.

## 1. Introduction

Magnetic van der Waals materials have emerged as promising building blocks for exploring fundamental low-dimensional magnetism and developing next-generation spintronic applications.[1–4] Among these, the transition-metal phosphorus trichalcogenide $NiPS_3$ has attracted significant interest due to its fascinating coherent exciton features, which are intimately coupled to its underlying zigzag antiferromagnetic order.[5] Extensive research has focused on the two-dimensional limit, where thinning the material down to a few layers has revealed unique characteristics in its magnetic phase transitions, significantly advancing our understanding of low-dimensional magnetic physics.[6–12] In this context, establishing a precise characterization of bulk properties is essential, especially using nearly twin-free samples, as it serves as a critical benchmark and counterpart to these 2D studies.

Despite its importance, experimentally resolving the intrinsic magnetic anisotropy of bulk twin-free $NiPS_3$ has often been challenging because bulk crystals are prone to crystallographic twinning.[13,14] These twins, manifesting as three distinct domains rotated by 120 ° relative to each other during stacking, effectively obscure the material's directional properties. In $NiPS_3$, the spin chain aligns along the monoclinic *a*-axis,[15] a direction governed by these twin domains. Consequently, the intrinsic anisotropic properties and any direction-dependent physical measurements of bulk $NiPS_3$ are often masked by structural twinning. We also note that the twinning issues have significant consequences for magnetic exciton physics and many other interesting properties reported for this material, as anisotropic characteristics are obscured or misinterpreted in their presence.[5,15–18]

In this study, we investigate the intrinsic magnetic and thermal properties of $NiPS_3$ using single crystals with minimal twinning, which effectively behave as single domains for bulk macroscopic measurements. We stress that twin-free crystals provide two major advantages: (i) they ensure high quality with minimum structural defects and (ii) they preserve the intrinsic directional information essential for precise angle-resolved measurements. By overcoming the limitations of twinning, we successfully resolve the intrinsic magnetic anisotropy and thermal transport properties, providing a comprehensive baseline for the bulk properties of $NiPS_3$.

## 2. Results

### 2.1. High-quality, nearly twin-free single crystal

To obtain high-quality, twin-free single crystals, we implemented a rigorous screening process that included X-ray diffraction (XRD), magnetization, and Raman spectroscopy. Our primary criterion for identifying and selecting the nearly twin-free single crystals heavily relied on the degree of macroscopic magnetization anisotropy. However, before this magnetic verification, we carried out a systematic pre-screening process. First, we carefully selected well-defined hexagonal-shaped crystals with flat and clean surfaces, ensuring there were no visible step edges or macroscopic terraces on the side facets. We then confirmed the high macroscopic crystalline quality of these pre-selected crystals using XRD, as shown in **Figure 1**(a).

From this narrowed-down pool of high-quality crystals, we measured the magnetic susceptibility ($\chi$) along the three candidate crystallographic a-axis directions. This step enabled us to rigorously evaluate in-plane magnetic anisotropy and ultimately isolate nearly twin-free samples. The presence of structural twins, primarily due to stacking faults, was assessed using anisotropy in magnetic susceptibility ($\chi$). Specifically, while twinned samples (Figure 1(b)) show mixed contributions from equivalent *a*-axes ($a_1$, $a_2$, and $a_3$), our selected twin-free crystals (Figure 1(c)) display clear anisotropic behavior with vanishingly small $\chi_a$ at base temperature.

Furthermore, for cross-validation, we performed polarized Raman spectroscopy on the top and bottom sides of the selected samples to independently reconfirm the consistency of the crystallographic *a*-axis. While magnetization probes the bulk-averaged structural and magnetic properties of the crystal, polarized Raman spectroscopy provides a convenient means to determine the crystallographic orientation at the surface. Due to the relatively shallow penetration depth of the 532 nm excitation laser (typically on the order of several tens to hundreds of nanometers), Raman spectroscopy primarily probes the near-surface region. Therefore, it serves as a sensitive tool for identifying surface twinning and determining the crystallographic orientation via the angular dependence of the $A_g$ phonon mode.

Figure 1(d) shows an unpolarized Raman spectrum measured at room temperature. Figures 1(e) and 1(f) present the angular dependence of the Raman intensity measured in the parallel linear polarization configuration using a 532 nm excitation laser. By correlating Raman results with magnetization measurements, we establish a correspondence between the polarization dependence of the $P_1$ mode and the crystallographic *a*-axis (indicated by the blue double-headed arrow).

In addition, we further quantified the portion of twin domains in our samples. Under the assumption that zero-temperature magnetic susceptibility along the *a*-axis approaches zero for a perfectly single-domain crystal, we estimate that the majority domain comprises roughly 86 % of the volume in our nearly twin-free samples (Figure 1(c); see Supporting Information for detailed calculations). For comparison, the twinned samples (Figure 1(b)) contain only 56 % of the majority domain.

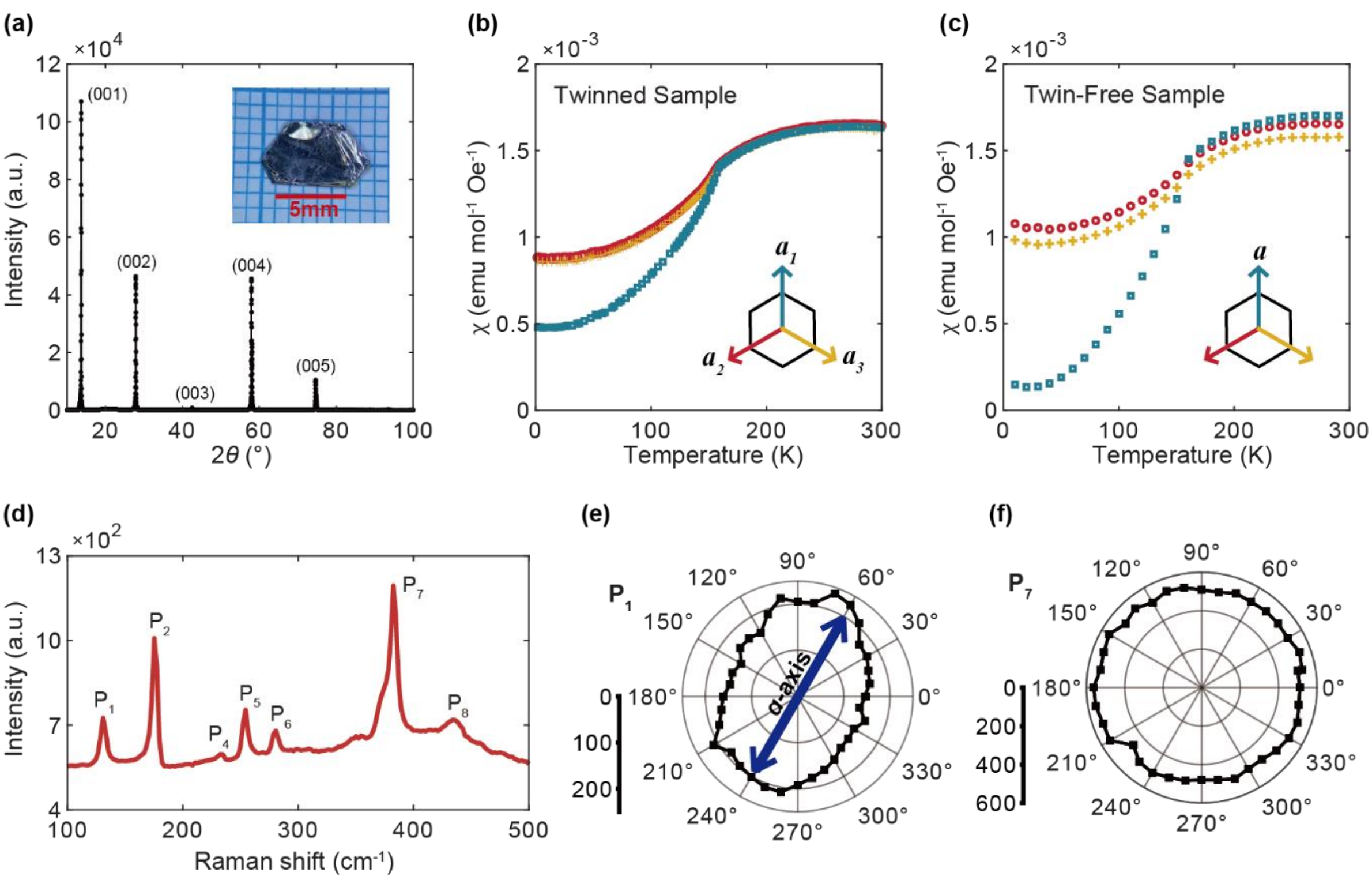

**Figure 1.** Structure and magnetic information of $NiPS_3$. (a) XRD pattern and image of a $NiPS_3$ single crystal. Magnetic susceptibility measured along three directions at 120° intervals from the a-axis for (b) a typical twinned crystal and (c) a crystal with minimal twinning. Identification of the a-axis on the surface using polarized Raman spectroscopy at room temperature. (d) Unpolarized Raman spectra of $NiPS_3$ at room temperature. (e,f) Angular dependence of the Raman intensity of the $P_1$ and $P_7$ phonon modes measured at room temperature using a 532nm excitation laser in the parallel linear polarization configuration.

## 2.2. Magnetization

As depicted in **Figure 2**(a), the magnetic ground state consists of zigzag chains with spins oriented predominantly along the *a*-axis. Consistent with this highly anisotropic arrangement, magnetic susceptibility measurements on our crystals exhibit a large anisotropy between the *a*- and *b*-axes below the transition temperature $T_N$~157.5 K (Figure 2(b)), confirming the twin-free nature and high quality of our crystals. Specifically, the magnetic susceptibility ratio ($\chi_{300K}/\chi_{2K}$) along the a-axis between 300 and 2 K is approximately 7.96. This value is significantly higher than the previously reported ratio of ~3,[15,19] indicating that our samples are highly untwinned. While achieving a perfectly pristine, macroscopic single domain remains technically prohibitive due to the thermodynamic nature of the stacking faults, our rigorous isolation process yields samples that effectively behave as single domains for bulk macroscopic measurements. In addition, our samples show a broad maximum in magnetic susceptibility above Néel temperature around 250 K due to short-range magnetic ordering, which suppresses the Curie-Weiss behavior until temperatures rise sufficiently to 400 K.[20,21]

Notably, a sharp spin-flop transition occurs when the magnetic field is applied along the a-axis, identifying the magnetic easy axis (see Figure 2(c)-(f)). At the base temperature, a metamagnetic transition is observed at approximately 10.5 T, which is consistent with previous reports.[8,22,23] The black open squares in Figure 2(c) and (d) represent the simulated magnetization for the monodomain structure at 1.8 K, using parameters adapted from A. Scheie *et al.*.[24] These simulation results are consistent with our observed magnetization and accurately capture the spin-flop transition. A comprehensive description of the simulation is provided in the Discussion and Experimental sections. Figure 2(e) illustrates the magnetic phase boundaries derived from the field derivative of magnetization ($dM/dH$). The peaks in $dM/dH$ allow us to map the temperature evolution of the spin-flop field ($H_{SF}$). The $H_{SF}$ increases as the temperature rises from the base temperature to $T_N$. We found that this spin reorientation persists up to 140 K within our magnetic field range.

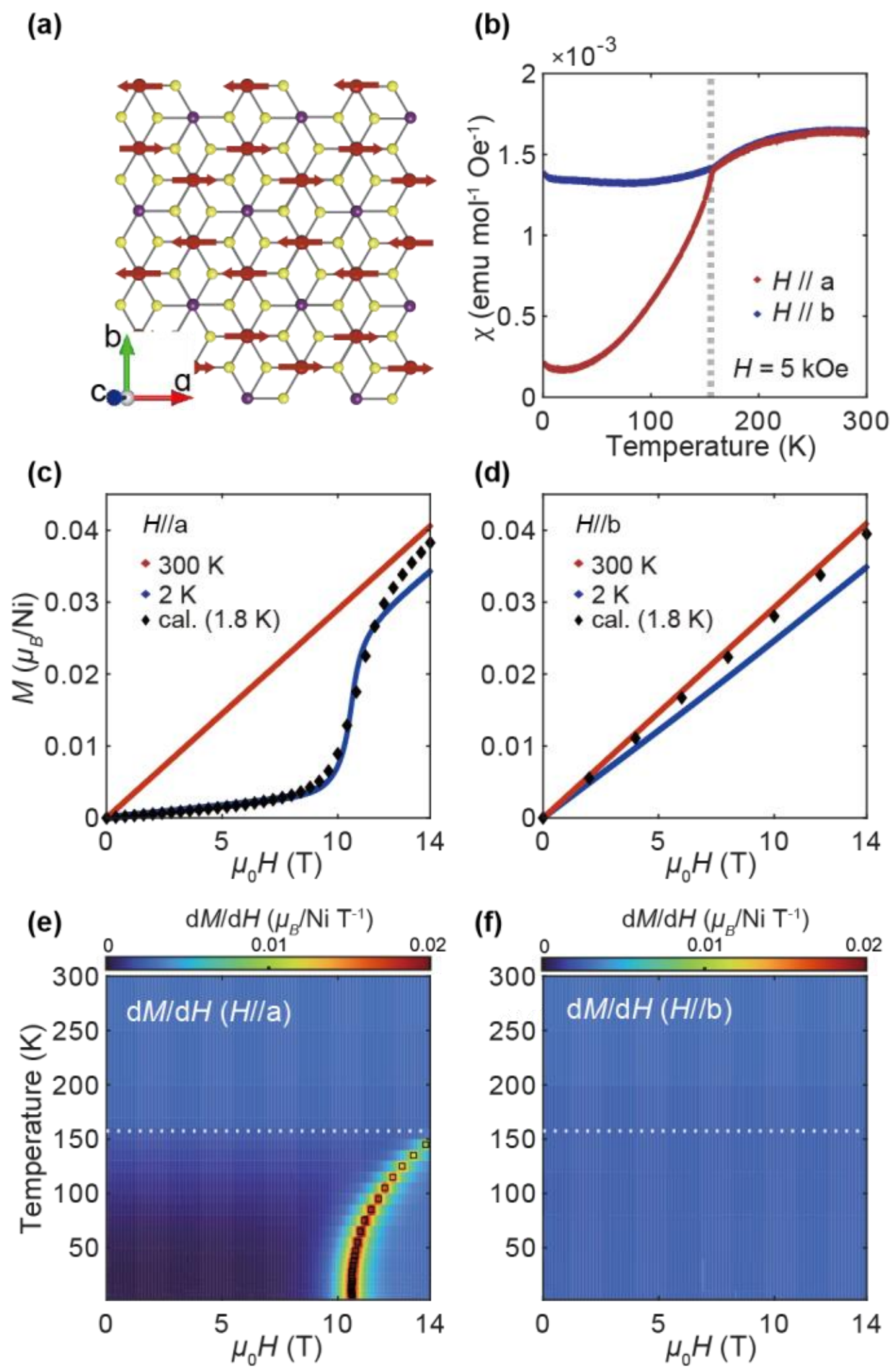

**Figure 2.** Magnetization measurement of a nearly twin-free $NiPS_3$ single crystal. (a) In-plane crystal and magnetic structures viewed along the *c**-axis. (b) Temperature dependence of magnetic susceptibility measured with a magnetic field of 5 kOe applied parallel to the *a*- and *b*-axes. Magnetic field dependence of magnetization along the (c) *a*-axis (d) *b*-axis. The calculated results for a perfect single-domain $NiPS_3$ crystal at 1.8 K are shown as black diamonds. (e)-(f) Contour plots of the differential magnetic susceptibility (*dM*/*dH*) as a function of temperature and magnetic field. White dotted lines indicate the antiferromagnetic phase transition temperature, and black open squares represent the *dM*/*dH* maxima at each measured temperature.

### 2.3. Specific heat

In **Figure 3**, the heat capacity data reveal a single lambda-like anomaly in the magnetic phase transition around 157 K. In particular, a broad magnetic contribution extends significantly around, which is a characteristic signature of quasi two-dimensional magnetic systems. Notably, as shown in the inset of Figure 3, the gradual and continuous recovery of magnetic entropy is observed above $T_N$. This slow release of magnetic entropy at temperatures well above $T_N$ is attributed to short-range low-dimensional spin fluctuations. While the magnetic entropy has not yet reached a full plateau within our experimental temperature range, its trajectory is approaching the theoretical limit of $Rln3$ expected for an $S$=1 system, corresponding to the $Ni^{2+}$ valence state of $NiPS_3$ in the paramagnetic phase.[21] Such behavior is consistent with the magnetic susceptibility observed above the transition temperature (Figure 2(b)).

To isolate this magnetic contribution, we estimated the lattice contribution on our heat capacity data employing a Debye-Einstein model. Previous studies on the phonon spectrum[25–28] reveal that the acoustic and low-lying optical branches extend up to ~ 35 meV. Additionally, distinct clusters of optical phonon modes are concentrated around 50 and 70 meV. Based on the energy distribution and spectral weight of these phonon branches, we established a model consisting of one Debye and three Einstein modes. Specifically, the Debye temperature ($\Theta_D$) was determined to be 201 K by fitting the low-temperature heat capacity data to a $T^3$ function. This combined model provided a reasonable fit to our experimental data (Table 1). By subtracting the estimated lattice contribution from the measured total heat capacity, we successfully extracted a magnetic contribution ($C_{mag}$) associated with the antiferromagnetic phase transition in $NiPS_3$. Subsequently, the temperature dependence of the magnetic entropy, $S_{mag}(T)$, was evaluated by integrating $C_{mag}/T$ over temperature as follows:

$$S_{mag}(T) = \int \frac{C_{mag}(T)}{T} dT \quad (1)$$

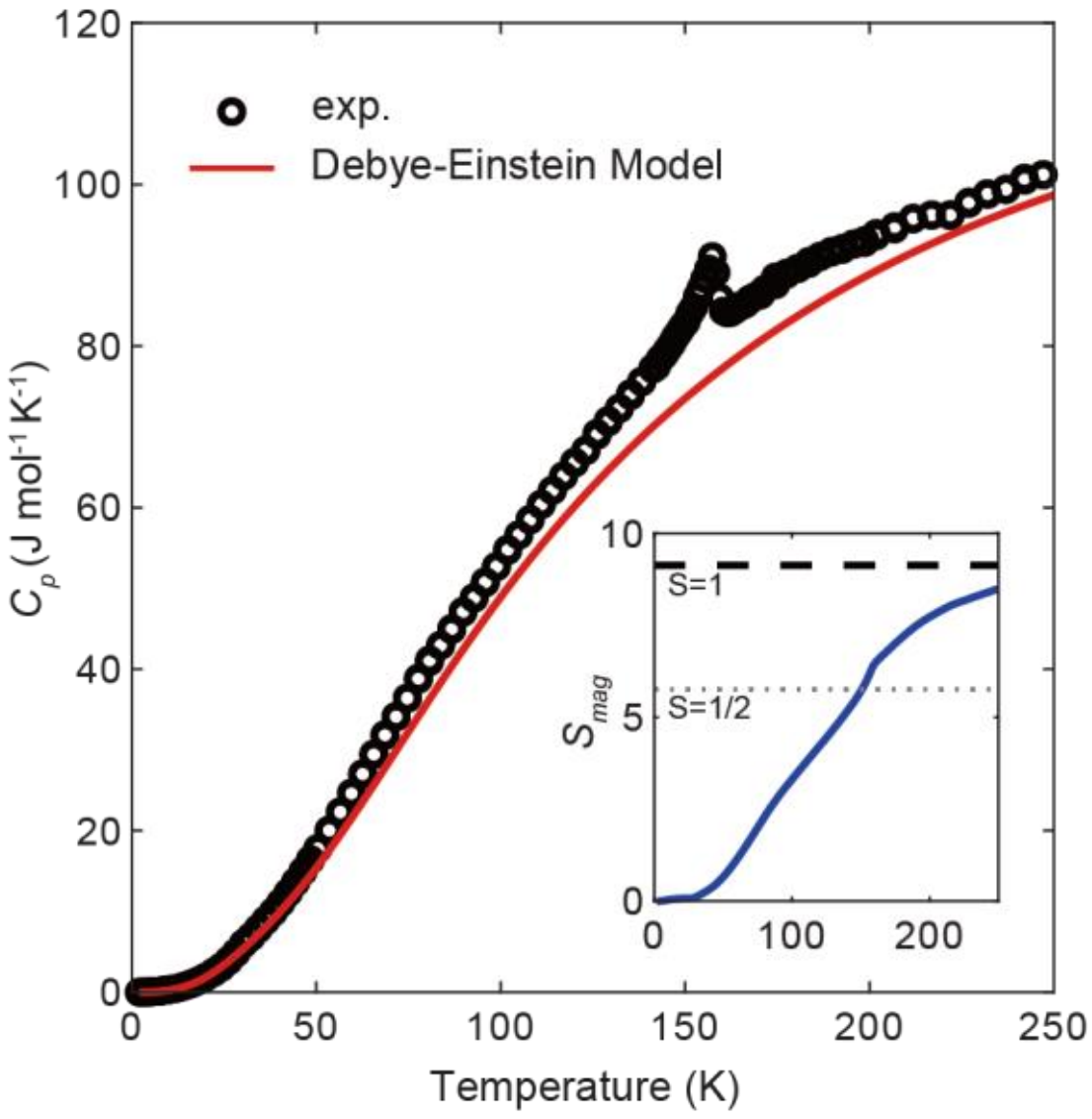

**Figure 3.** Temperature dependence of heat capacity per formula unit. The phonon contribution (red solid line) was estimated using the Debye-Einstein model. The inset shows the magnetic entropy derived from the magnetic contribution.

**Table 1.** Parameters of the Debye-Einstein model.

| | $\Theta_D$ | $\Theta_{E1}$ | $\Theta_{E2}$ | $\Theta_{E3}$ | Total |
|---|---|---|---|---|---|
| Temperature [K] | 201 | 355 | 580 | 810 | |
| Weight | 3R | 8.5R | 1R | 2.5R | 15R |

## 2.4. Thermal transport

**Figure 4** displays thermal conductivity ($\kappa$) measured along the $a$-axis. Figure 4(a) shows the temperature dependence of $\kappa$ in the absence of a magnetic field, fitted using the Debye-Callaway model[29,30] to estimate a non-magnetic baseline. $\kappa$ peaks at 22 K with a maximum magnitude of 374.5 W $K^{-1}$ $m^{-1}$. Notably, through a wide temperature range around $T_N$, the thermal conductivity deviates significantly from the typical $1/T$ dependence expected for dominant Umklapp scattering.[29] This suppression suggests a substantial scattering of phonons by spin fluctuations.[31–34]

Furthermore, Figures 4(b) and 4(c) present the field-dependent thermal conductivity measurements. The magnetic field was applied along the $a$-axis, parallel to the in-plane spin alignment. Despite a subtle field dependence (changes limited to ~4%) in comparison to materials exhibiting dramatic spin-lattice effects,[35–40] the thermal conductivity remains highly sensitive to the spin-flop transition. Below 20 K, the spin-flop transition induces a sharp dip in $\kappa$. Even at higher temperatures, traces of the transition are discernible in the field derivative ($d\kappa/dH$). These features coincide with the phase boundary determined from the maxima in $dM/dH$. Beyond the vicinity of the spin-flop transition, the thermal conductivity exhibits a variety of field-dependent behaviors across different temperature regimes. At 6 K, the thermal conductivity decreases gradually with increasing field and then recovers sharply immediately after the transition. Near 10 K, the field dependence remains relatively weak, except for a characteristic dip at the $H_{SF}$. Interestingly, between 11 and 18 K, the thermal conductivity shows a positive dependence on the magnetic field, while the recovery of $\kappa$ following the spin-flop transition progressively weakens. As the temperature increases further, these field-dependent features gradually diminish.

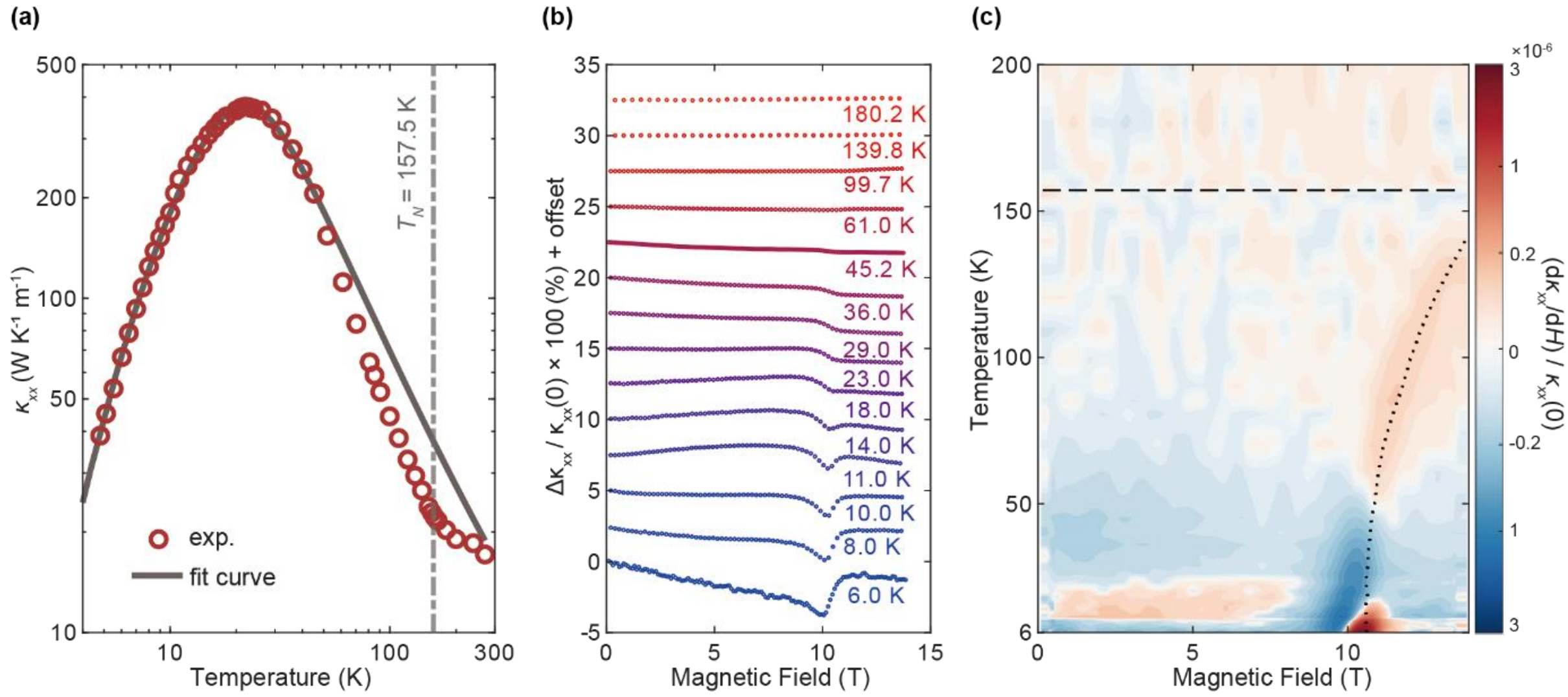

**Figure 4.** Thermal transport properties along the *a*-axis. (a) Temperature dependence of the thermal conductivity. The solid line represents a fit using the Debye-Callaway model. (b) Magnetic field dependence of the normalized thermal conductivity changes at various temperatures (T = 6, 8, 10, 11, 14, 18, 23, 29, 36, 45.2, 61, 99.7, 139.8, and 180.2 K). (c) Contour plot of the field derivative of the normalized thermal conductivity. The dashed and dotted lines indicate the antiferromagnetic phase transition temperatures and the spin-flop transitions, respectively, determined from magnetization data.

## 3. Discussion

### 3.1. Debye-Callaway model fitting and spin fluctuation around $T_N$

In Figure 4(a), we present the thermal conductivity of the lattice background using the Debye-Callaway model.[29,30] The relatively small magnetic field dependence suggests that phonons are the dominant heat carriers in $NiPS_3$. To quantitatively describe this lattice background, we employed the Debye-Callaway model:

$$\kappa = \frac{k_B}{2\pi^2 v_D}\left(\frac{k_B}{\hbar}\right)^3 T^3 \int_0^{\frac{\Theta_D}{T}} \tau \frac{x^4 e^x}{(e^x-1)^2} dx \tag{2}$$

where $k_B$ is the Boltzmann constant, $\hbar$ is the reduced Planck constant, $\omega$ represents the angular frequency of phonons, $T$ is the temperature, and $x=\hbar\omega/k_BT$. The Debye temperature ($\Theta_D$) and the Debye velocity ($v_D$) were adopted from our specific heat analysis and represent the acoustic phonon characteristics that primarily govern thermal transport. Here, $v_D$ was determined via the relation $\Theta_D = v_D \frac{\hbar}{k_B}(6\pi^2 n)^{1/3}$, where $n$ denotes 1/(volume of formula unit). The total phonon scattering rate, $\tau^{-1}$, was established using Matthiessen's rule, accounting for various scattering mechanisms:

$$\tau^{-1} = \tau_{\mathrm{BD}}^{-1} + \tau_{\mathrm{LD}}^{-1} + \tau_{\mathrm{PD}}^{-1} + \tau_{\mathrm{U}}^{-1} \tag{3}$$

This scattering rate incorporates Boundary scattering ($\tau_{\mathrm{BD}}$) (4), line defects ($\tau_{\mathrm{LD}}$) (5), point defects ($\tau_{\mathrm{PD}}$) (6), and Umklapp scattering ($\tau_{\mathrm{U}}$) (7)[29,37–39,41,42]. By fitting the experimental data to this scattering model, we established a robust lattice background with the parameters listed in Table 2.

$$\tau_{\mathrm{BD}}^{-1} = \frac{v_{\mathrm{D}}}{d} \tag{4}$$

$$\tau_{\mathrm{LD}}^{-1} = A_0 \omega \tag{5}$$

$$\tau_{\mathrm{PD}}^{-1} = A_1 \omega^4 \tag{6}$$

$$\tau_{\mathrm{U}}^{-1} = A_2 \omega^2 T exp(-\frac{\Theta_{\mathrm{D}}}{bT}) \tag{7}$$

Where $d$ represents the characteristic sample size, $b$ is an order-of-unity parameter for the umklapp processes, and $A_0$, $A_1$, and $A_2$ are free parameters. In comparing experimental data with model fits, it is notable that dimensionality-enhanced spin fluctuations appear to provide an additional scattering mechanism near the antiferromagnetic phase transition. Although one might expect additional scattering from spin fluctuations to exhibit field dependence, we observed negligible changes in $\kappa$ near $T_{\mathrm{N}}$ up to 14 T. This lack of field dependence can be understood in the following manner: the representative energy scale of the exchange interaction is the Curie-Weiss temperature ($\Theta_{\mathrm{CW}} \approx -250$ K)[21], corresponding to an effective magnetic field of ~186 T. Thus, this intrinsic energy scale is significantly larger than the magnetic-field energy within our 14 T range, effectively masking any field-induced changes to this scattering channel. Consequently, heat transport in this system is significantly governed by phonon scattering from spin fluctuations.

**Table 2.** Parameters of the Debye-Callaway thermal conductivity model.

| Parameters [units] | Value |
|---|---|
| $\Theta_D$ [K] | 201 |
| $v_D$ $[m/s]$ | 3057 |
| $d$ $[mm]$ | 0.1 |
| $A_0$ $[unitless]$ | $1.1745 \times 10^{-7}$ |
| $A_1$ $[s^3]$ | $5.0967 \times 10^{-44}$ |
| $A_2$ $[s\ K^{-1}]$ | $1.4631 \times 10^{-18}$ |
| $b$ $[unitless]$ | 2.3945 |

### 3.2. Spin wave calculation and Spin-flop transition

To further investigate the spin-flop transition in $NiPS_3$, we calculated the spin-wave dispersion and the field-dependent magnon gap using the SpinW package.[43] The spin Hamiltonian was adapted from previous neutron spectroscopy studies,[24] incorporating a canted spin arrangement in which spins are tilted approximately 13.5 ° out of the *ab*-plane from the *a*-axis.[15] The Hamiltonian is defined as:

$$\mathcal{H} = \sum_{i\,j} J_{ij}\boldsymbol{S}_i \cdot \boldsymbol{S}_j + \sum_i [A_{\tilde{x}}(S_i^{\tilde{x}})^2 + A_{\tilde{z}}(S_i^{\tilde{z}})^2] - g\mu_{\mathrm{B}}\boldsymbol{B} \cdot \sum_i \boldsymbol{S}_i \quad (8)$$

where $J_{ij}$ represents the exchange interaction strength, $A_{\tilde{x}}$ and $A_{\tilde{z}}$ are the single-ion anisotropy terms, $g$ is the $g$-factor (taken as $g$=2), $\mu_{\mathrm{B}}$ is the Bohr magneton, and $\boldsymbol{B}$ is the external magnetic field. We utilized the exchange parameters from previous reports,[24] including four in-plane interactions ($J_{1\mathrm{a}}$, $J_{1\mathrm{b}}$, $J_2$, and $J_3$) and one out-of-plane interaction ($J_4$) (**Figure 5**(a)).

The single-ion anisotropy parameters obtained from the inelastic neutron scattering results of Scheie *et al.*[24] are $A_x$= -0.010±0.005 meV and $A_z$= 0.21±0.03 meV. In our model, we used $A_{\tilde{x}}$ = -0.009 meV and $A_{\tilde{z}}$ = 0.21 meV to reproduce the spin-flop transition around 10.5 T, with $A_{\tilde{x}}$ still lying within the reported uncertainty. Furthermore, the rotation of the anisotropy axes by $\theta_{\mathrm{A}}$=13.5°, $\tilde{x} = \cos\theta_{\mathrm{A}}\,\hat{x} - \sin\theta_{\mathrm{A}}\,\hat{z}, \tilde{z} = \sin\theta_{\mathrm{A}}\,\hat{x} + \cos\theta_{\mathrm{A}}\,\hat{z}$ (Figure 5(b)), which reflects the experimentally reported canting of the ordered moment [15]. Based on the proposed spin Hamiltonian, we simulated the field-dependent magnetization at 1.8 K, as shown in Figure 2(c) and 2(d). The simulation results show excellent agreement with the measured data, particularly in capturing the sharp increase in the *a*-axis magnetization associated with the spin-flop transition.

We also note that the small rotation in anisotropy axes does not influence the zero-field magnon dispersion; it is crucial for obtaining the correct single-domain field response for *H* || *a*. After incorporating the canting angle measured by neutron diffraction into the Hamiltonian, the single-domain calculation is taken as the expected *M*(*H*) response of the twin-free limit. The close agreement between our measured curves and this single-domain response, including the sharp spin-flop transition for *H* || *a* and the absence of a corresponding transition for *H* || *b*, supports that the selected crystals are nearly twin-free.

**Figure 6**(a) displays spin-wave dispersions at *B* = 0, 10.75, and 20 T. Upon applying a field, the magnon energy levels undergo Zeeman splitting, which closes the magnon gap at the zone center until the spin-flop transition. We tracked the evolution of the magnon gap as a function of magnetic field (Figure 6(b)), observing that the gap narrows to approximately 0.07 meV at the transition point (10.75 T) and then recovers in the spin-flopped phase.

The relationship between the magnon gap and thermal conductivity is critical for understanding the characteristic dip observed at $H_{\mathrm{SF}}$. We introduced a simple phonon-scattering model in which the scattering rate is proportional to the magnon population, estimated from the calculated magnon gap. Specifically, the additional scattering rate $\tau_{\text{mag}}^{-1}$ is given by:

$$\tau_{\text{mag}}^{-1} = A_{\text{mag}}[\exp(\Delta/k_{\text{B}}T) - 1]^{-1} \quad (9)$$

where Δ denotes the calculated magnon gap energy. As shown in Figure 6(c), this additional scattering mechanism—with a representative value of $A_{\text{mag}}=1.5\times10^5$ and T = 10 K adopted to illustrate the qualitative trend—effectively reproduces the characteristic dip in $\kappa$ at the spin-flop transition.

Nevertheless, the field dependence of $\kappa$ exhibits a non-trivial evolution across different temperature regimes, suggesting that additional competing spin-lattice coupling mechanisms are involved in the heat transport of $NiPS_3$. These features may be tentatively attributed to two primary factors: not only the scattering source, but also the role of magnons as heat carriers[44,45] and the involvement of magnon-phonon hybridization.[19,38,42,46,47] In the former case, a field-induced reduction of the magnon gap increases the population of magnonic excitations, which could enhance the total thermal conductivity. Concurrently, hybridization may occur at the crossing points between magnon and acoustic phonon branches. In Figure 6(d), we compared the computed phonon spectrum of $NiPS_3$[28] with our magnon spectrum calculations. We reveal crossings between the dispersions along Γ-Y-M', with the lowest-energy crossing occurring near the acoustic zone boundary (~9 meV).[19] This intersection may serve as an additional scattering source, offering a plausible explanation for the anomalous features that a simple gap-driven model cannot fully account for. While various transport and scattering processes remain at play, a definitive identification of their mechanisms requires further investigation.

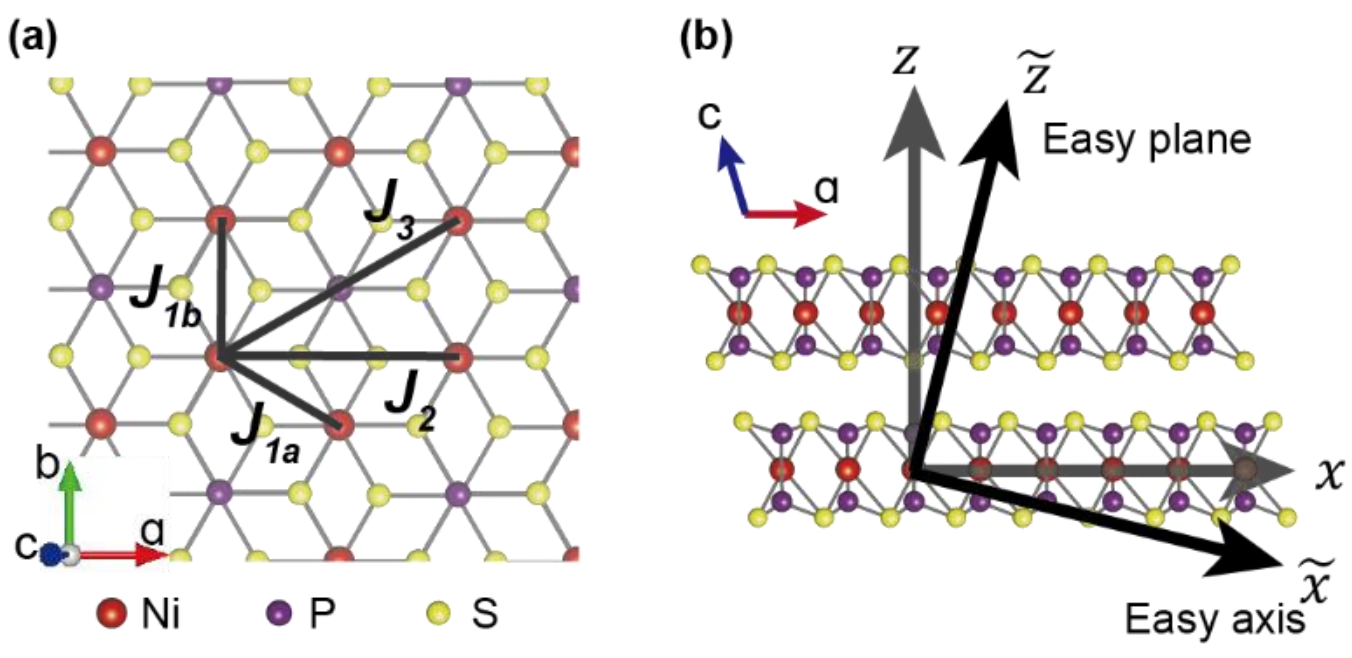

**Figure 5.** (a) Schematic illustration of the spin Hamiltonian parameters, showing the intra-layer exchange interactions ($J_{1a}$, $J_{1b}$, $J_2$, and $J_3$). (b) Orientation of the single-ion anisotropy

axes used in the spin Hamiltonian. We defined $\tilde{x}$ and $\tilde{z}$ as the new easy axis and easy plane, respectively, instead of $x$ and $z$.

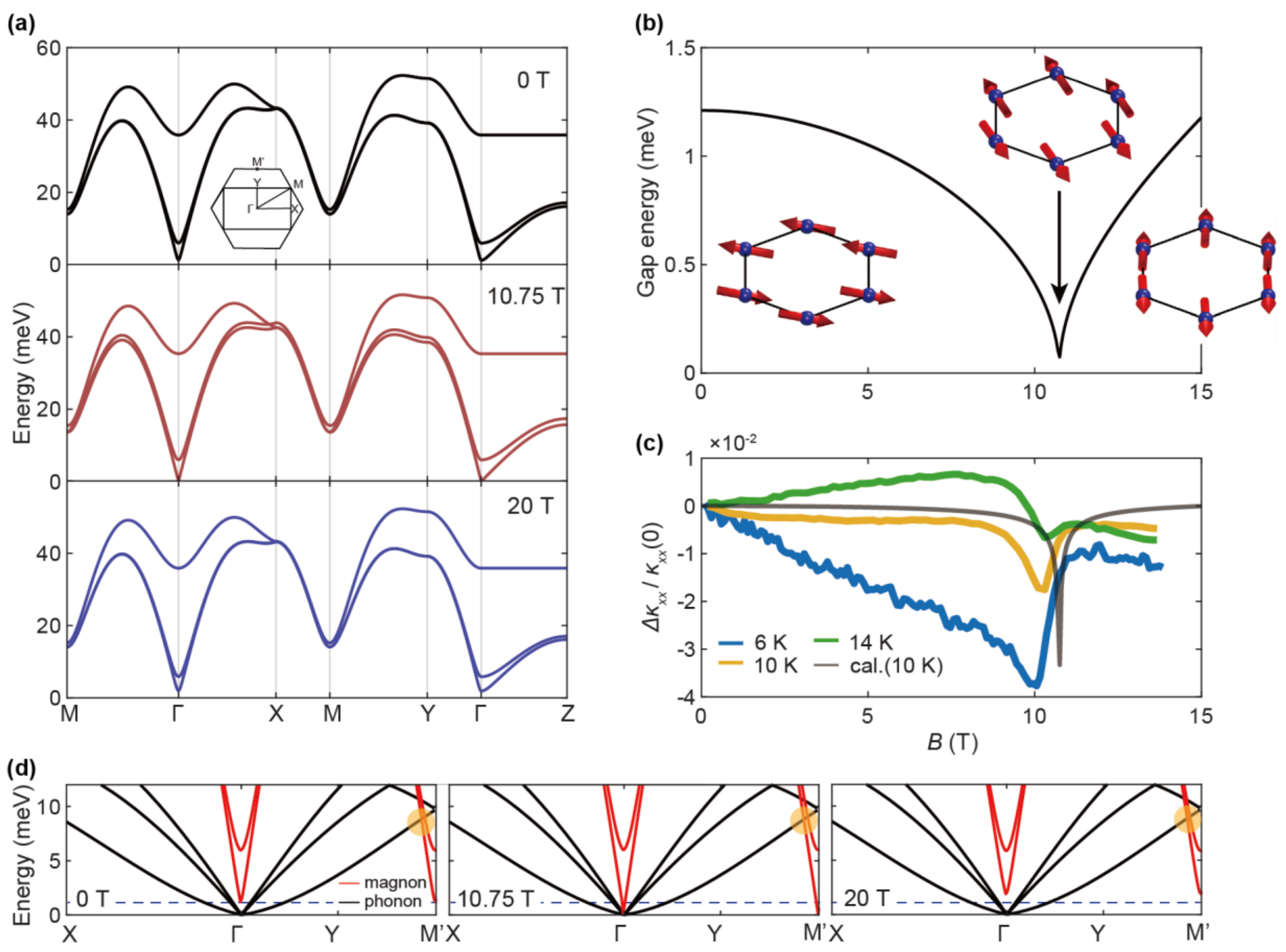

**Figure 6.** (a) Spin wave calculation under an external magnetic field applied along the a-axis at $B$ = 0, 10.5, and 20 T. (b) Calculated magnon gap size as a function of the magnetic field and corresponding magnetic ground states for each phase. (c) Comparison of experimental thermal conductivities (colored lines) with simulated results (grey line). (d) Calculated low-energy magnon and phonon[28] spectrum. The orange points indicate the lowest crossing points between the magnon and phonon branches. To make the field dependence visible in the dispersion in (e), we added the horizontal blue dashed line corresponding to the energy at the Γ point at 0 T.

## 4. Conclusion

In summary, we have provided a comprehensive thermodynamic study of nearly twin-free $NiPS_3$ single crystals, including measurements of intrinsic anisotropy, magnetization, heat capacity, and thermal transport. Our rigorous screening process ensured the selection of twin-free samples from the synthesized batches, yielding a magnetic susceptibility ratio ($\chi_{300K}/\chi_{2K}$~7.96) that is significantly higher than previously reported data. The analysis of heat capacity and thermal conductivity demonstrates that spin fluctuations exert a significant influence over a broad temperature range around $T_N$, a characteristic behavior of low-dimensional magnetism. Additionally, the observation of a sharp spin-flop transition in field-dependent thermal conductivity provides evidence of spin-lattice coupling, which acts as a scattering source during the spin reorientation. Lastly, through spin-wave calculations, we further elucidated the evolution of the field-dependent magnon gap. We successfully modelled the associated transport anomalies, which effectively capture the characteristic features observed at the transition.

Our work demonstrates that identifying nearly twin-free samples is feasible and essential for resolving intrinsic anisotropies, suggesting that crystallographic twinning can and must be carefully addressed in future direction-dependent studies of van der Waals magnets. Using these high-quality crystals, we provide a definitive baseline for understanding the thermodynamic properties of bulk twin-free $NiPS_3$. Crucially, the robust spin-lattice coupling and thermal transport anomalies characterized here provide a necessary foundation for engineering thermal and magnetic functionalities in future low-dimensional spintronic architectures.

## 5. Experimental Section

Single crystals were synthesized via a chemical vapor transport (CVT) method from high-purity powdered elemental sources. Ni (>99.99%), P (>99.99%), and S (>99.998%) powders were mixed in a 1:1:3 molar ratio inside an Ar-filled glove box. For vapor transport, we added 5% sulfur to the mixture.[24] Structural characterization was conducted using X-ray diffraction (XRD) (MINIFLEX II, Rigaku) and Raman spectroscopy (XperRam Compact, Nanobase). The elemental composition of the crystals was analyzed by energy-dispersive X-ray spectroscopy (EDX) (Quantax 100, Bruker, and EM-30, COXEM), confirming a representative composition of $Ni_{0.97}PS_{2.94}$.

Subsequently, high-quality samples with minimal twinning were selected based on magnetic susceptibility measurements, along with other supporting methodologies developed for this work (see Section 2.1). Magnetization measurements were performed on the single-crystal samples using commercial magnetometers (PPMS Dynacool VSM and MPMS-XL5, Quantum Design). Specific heat was measured using a Physical Property Measurement System (PPMS, Quantum Design, USA). Thermal conductivity was determined using the steady-state method in a 14 T magnet cryostat (CFMS14T, Cryogenic Ltd., UK). The thermal transport setup was custom-made, employing a single heater and two thermometers. We used homemade $SrTiO_3$ capacitance thermometers that exhibit excellent accuracy under high magnetic fields.[48] To minimize geometric effects, high-quality single crystals were prepared in a rectangular strip shape ($4.4 \times 0.52 \times 0.22$ mm$^3$) with the longest dimension aligned along the *a*-axis.

For our theoretical studies, we used a spin Hamiltonian with parameters reported from two inelastic neutron scattering experiments,[13,24] with a few adjustments to fit our data. For example, magnetization simulations were conducted at 1.8 K using the SUNNY package,[49] employing the Monte Carlo method and a Landau-Lifshitz dynamics (LLD) sampler with a time step of $dt$=0.00719 meV$^{-1}$, and a damping parameter $\lambda$=0.1. A 6×6×2 supercell of a monoclinic unit cell was used to calculate the field-dependent magnetization.

**Acknowledgements**

We thank Beom Hyun Kim, and Chaebin Kim for valuable discussions. This work was supported by the Leading Researcher Program of the National Research Foundation of Korea (Grant No. RS-2020-NR049405) & the Core Center Program (2021R1A6C101B418), the Ministry of Education of the Korean government.

**Data Availability Statement**

The data that support the findings of this study are available from the corresponding author upon reasonable request.

**References**

[1] J.-G. Park, *Journal of Physics: Condensed Matter* **2016**, *28*, 301001.
[2] K. S. Burch, D. Mandrus, J.-G. Park, *Nature* **2018**, *563*, 47.
[3] J.-G. Park, K.-X. Zhang, H. Cheong, J. H. Kim, C. A. Belvin, D. Hsieh, H. Ning, N. Gedik, *Rev. Mod. Phys.* **2026**, *98*, 025003.
[4] D. L. Cortie, G. L. Causer, K. C. Rule, H. Fritzsche, W. Kreuzpaintner, F. Klose, *Adv. Funct. Mater.* **2020**, *30*, 1901414.
[5] S. Kang, K. Kim, B. H. Kim, J. Kim, K. I. Sim, J.-U. Lee, S. Lee, K. Park, S. Yun, T. Kim, A. Nag, A. Walters, M. Garcia-Fernandez, J. Li, L. Chapon, K.-J. Zhou, Y.-W. Son, J. H. Kim, H. Cheong, J.-G. Park, *Nature* **2020**, *583*, 785.

[6] K. Kim, S. Y. Lim, J.-U. Lee, S. Lee, T. Y. Kim, K. Park, G. S. Jeon, C.-H. Park, J.-G. Park, H. Cheong, *Nat. Commun.* **2019**, *10*, 345.
[7] Z. Sun, G. Ye, C. Zhou, M. Huang, N. Huang, X. Xu, Q. Li, G. Zheng, Z. Ye, C. Nnokwe, L. Li, H. Deng, L. Yang, D. Mandrus, Z. Y. Meng, K. Sun, C. R. Du, R. He, L. Zhao, *Nat. Phys.* **2024**, *20*, 1764.
[8] C.-Y. Cheon, V. Multian, K. Watanabe, T. Taniguchi, A. F. Morpurgo, D. Lebedev, *Nat. Commun.* **2025**, *17*, 60.
[9] F. Y. Gao, D. S. Kim, C. Lei, A. Kumar, X. Peng, X. Liu, F. Barantani, S. Zhang, K. P. Lee, K. Raju, D. Lujan, S. Arash, S. Raman, S.-F. Lee, M. Ye, X. Li, A. H. MacDonald, E. Baldini, *Nat. Mater.* **2026**, 1.
[10] C.-T. Kuo, M. Neumann, K. Balamurugan, H. J. Park, S. Kang, H. W. Shiu, J. H. Kang, B. H. Hong, M. Han, T. W. Noh, J.-G. Park, *Sci. Rep.* **2016**, *6*, 20904.
[11] W. Na, P. Park, S. Oh, J. Kim, A. Scheie, D. A. Tennant, H. C. Lee, J.-G. Park, H. Cheong, *ACS Nano* **2024**, *18*, 20482.
[12] F. Wang, T. A. Shifa, P. Yu, P. He, Y. Liu, F. Wang, Z. Wang, X. Zhan, X. Lou, F. Xia, J. He, *Adv. Funct. Mater.* **2018**, *28*, 1802151.
[13] A. R. Wildes, J. R. Stewart, M. D. Le, R. A. Ewings, K. C. Rule, G. Deng, K. Anand, *Phys. Rev. B* **2022**, *106*, 174422.
[14] C. Murayama, M. Okabe, D. Urushihara, T. Asaka, K. Fukuda, M. Isobe, K. Yamamoto, Y. Matsushita, *J. Appl. Phys.* **2016**, *120*, 142114.
[15] A. R. Wildes, V. Simonet, E. Ressouche, G. J. McIntyre, M. Avdeev, E. Suard, S. A. J. Kimber, D. Lançon, G. Pepe, B. Moubaraki, T. J. Hicks, *Phys. Rev. B* **2015**, *92*, 224408.
[16] X. Wang, J. Cao, Z. Lu, A. Cohen, H. Kitadai, T. Li, Q. Tan, M. Wilson, C. H. Lui, D. Smirnov, S. Sharifzadeh, X. Ling, *Nat. Mater.* **2021**, *20*, 964.
[17] D. S. Kim, D. Huang, C. Guo, K. Li, D. Rocca, F. Y. Gao, J. Choe, D. Lujan, T. Wu, K. Lin, E. Baldini, L. Yang, S. Sharma, R. Kalaivanan, R. Sankar, S. Lee, Y. Ping, X. Li, *Advanced Materials* **2023**, *35*, 2206585.
[18] F. Song, Y. Lv, Y.-J. Sun, S. Pang, H. Chang, S. Guan, J.-M. Lai, X.-J. Wang, B. Wu, C. Hu, Z. Yuan, J. Zhang, *Nat. Commun.* **2024**, *15*, 7841.
[19] Q. Meng, X. Li, J. Liu, L. Zhao, C. Dong, Z. Zhu, L. Li, K. Behnia, **2024**, *arXiv:2403.13306*.
[20] S. Y. Kim, T. Y. Kim, L. J. Sandilands, S. Sinn, M.-C. Lee, J. Son, S. Lee, K.-Y. Choi, W. Kim, B.-G. Park, C. Jeon, H.-D. Kim, C.-H. Park, J.-G. Park, S. J. Moon, T. W. Noh, *Phys. Rev. Lett.* **2018**, *120*, 136402.
[21] P. A. Joy, S. Vasudevan, *Phys. Rev. B* **1992**, *46*, 5425.
[22] D. Jana, P. Kapuscinski, I. Mohelsky, D. Vaclavkova, I. Breslavetz, M. Orlita, C. Faugeras, M. Potemski, *Phys. Rev. B* **2023**, *108*, 115149.
[23] X. Wang, J. Cao, Z. Lu, A. Cohen, H. Kitadai, T. Li, Q. Tan, M. Wilson, C. H. Lui, D. Smirnov, S. Sharifzadeh, X. Ling, *Nat. Mater.* **2021**, *20*, 964.
[24] A. Scheie, P. Park, J. W. Villanova, G. E. Granroth, C. L. Sarkis, H. Zhang, M. B. Stone, J.-G. Park, S. Okamoto, T. Berlijn, D. A. Tennant, *Phys. Rev. B* **2023**, *108*, 104402.
[25] A. Hashemi, H.-P. Komsa, M. Puska, A. V. Krasheninnikov, *The Journal of Physical Chemistry C* **2017**, *121*, 27207.
[26] Z. Muhammad, J. Szpakowski, G. Abbas, L. Zu, R. Islam, Y. Wang, F. Wali, A. Karmakar, M. R. Molas, Y. Zhang, L. Zhu, W. Zhao, H. Zhang, *2d Mater.* **2023**, *10*, 025001.
[27] J. Yang, X. Zhang, Z. Qu, Q. Min, J. Xia, H. Shang, S. Luo, C. Wu, *RSC Adv.* **2025**, *15*, 23115.
[28] Y. Liu, Y. Liu, J. Zhao, X. Jiang, *ACS Appl. Nano Mater.* **2025**, *8*, 2291.
[29] R. Berman, *Thermal conduction in solids*, Clarendon, Oxford, **1978**.
[30] J. Callaway, *Physical Review* **1959**, *113*, 1046.
[31] G. A. Slack, *Physical Review* **1961**, *122*, 1451.

[32] G. A. Slack, R. Newman, *Phys. Rev. Lett.* **1958**, *1*, 359.
[33] D. D. Vu, R. A. Nelson, B. L. Wooten, J. Barker, J. E. Goldberger, J. P. Heremans, *Phys. Rev. B* **2023**, *108*, 144402.
[34] K. Yang, H. Wu, Z. Li, C. Ran, X. Wang, F. Zhu, X. Gong, Y. Liu, G. Wang, L. Zhang, X. Mi, A. Wang, Y. Chai, Y. Su, W. Wang, M. He, X. Yang, X. Zhou, *Adv. Funct. Mater.* **2023**, *33*, 2302191.
[35] T. Ideue, T. Kurumaji, S. Ishiwata, Y. Tokura, *Nat. Mater.* **2017**, *16*, 797.
[36] H. Yang, X. Xu, J. H. Lee, Y. S. Oh, S.-W. Cheong, J.-G. Park, *Phys. Rev. B* **2022**, *106*, 144417.
[37] H. Yang, C. Kim, Y. Choi, J. H. Lee, G. Lin, J. Ma, M. Kratochvílová, P. Proschek, E.-G. Moon, K. H. Lee, Y. S. Oh, J.-G. Park, *Phys. Rev. B* **2022**, *106*, L081116.
[38] H. Yang, G. Go, J. Park, S. K. Kim, J.-G. Park, *Phys. Rev. B* **2024**, *110*, 165147.
[39] P. A. Sharma, J. S. Ahn, N. Hur, S. Park, S. B. Kim, S. Lee, J.-G. Park, S. Guha, S.-W. Cheong, *Phys. Rev. Lett.* **2004**, *93*, 177202.
[40] C. Xu, C. Carnahan, H. Zhang, M. Sretenovic, P. Zhang, D. Xiao, X. Ke, *Phys. Rev. B* **2023**, *107*, L060404.
[41] A. V. Sologubenko, K. Giannó, H. R. Ott, U. Ammerahl, A. Revcolevschi, *Phys. Rev. Lett.* **2000**, *84*, 2714.
[42] S. Bao, Z. Cai, W. Si, W. Wang, X. Wang, Y. Shangguan, Z. Ma, Z.-Y. Dong, R. Kajimoto, K. Ikeuchi, S.-L. Yu, J. Sun, J.-X. Li, J. Wen, *Phys. Rev. B* **2020**, *101*, 214419.
[43] S. Toth, B. Lake, *Journal of Physics: Condensed Matter* **2015**, *27*, 166002.
[44] C. Hess, C. Baumann, U. Ammerahl, B. Büchner, F. Heidrich-Meisner, W. Brenig, A. Revcolevschi, *Phys. Rev. B* **2001**, *64*, 184305.
[45] S. R. Boona, J. P. Heremans, *Phys. Rev. B* **2014**, *90*, 064421.
[46] G. Laurence, D. Petitgrand, *Phys. Rev. B* **1973**, *8*, 2130.
[47] M. Nawwar, R. R. Neumann, J. Wen, I. Mertig, A. Mook, J. P. Heremans, *Reports on Progress in Physics* **2025**, *88*, 080503.
[48] H.-L. Kim, M. J. Coak, J. C. Baglo, K. Murphy, R. W. Hill, M. Sutherland, M. C. Hatnean, G. Balakrishnan, J.-G. Park, *Review of Scientific Instruments* **2019**, *90*.
[49] D. Dahlbom, H. Zhang, C. Miles, S. Quinn, A. Niraula, B. Thipe, M. Wilson, S. Matin, H. Mankad, S. Hahn, D. Pajerowski, S. Johnston, Z. Wang, H. Lane, Y. W. Li, X. Bai, M. Mourigal, C. D. Batista, K. Barros, *J. Open Source Softw.* **2025**, *10*, 8138.

**Supporting Information: Probing intrinsic magnetic phases in low-dimensional nearly twin-free $NiPS_3$ single crystals**

*Yeochan An#, Heejun Yang#, Sung Jin Park#, Giung Park#, Woonghee Cho, Pyeongjae Park, Seokhwan Yun, Yoshimitsu Kohama, and Je-Geun Park**

## 1. Estimating the domain portion

We estimate the volume fraction of each domain using our magnetic susceptibility data with three different $a$-axes. Assuming that magnetic susceptibility has negligible off-diagonal components and $\chi_{aa}$, then we can get in-plane magnetic susceptibility with arbitrary angle $\phi$.

$$\chi_\phi = \chi_{bb} \sin^2 \phi \quad \text{(s1)}$$

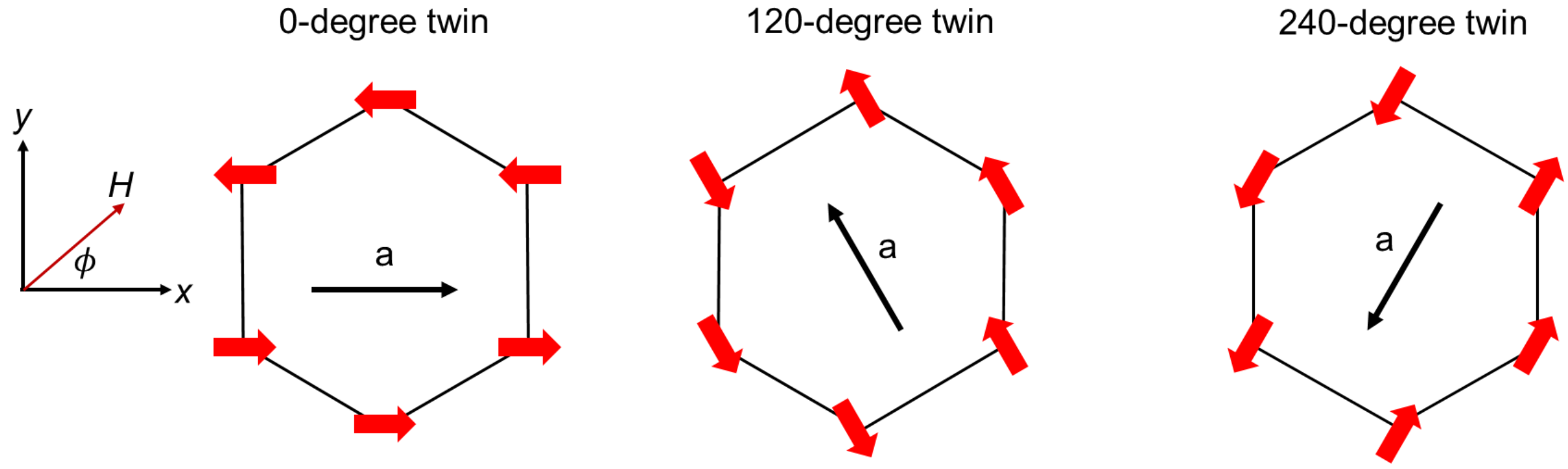

**Figure S1**. Schematic for three twin domains

We can express the magnetization in the following manner for a sample with 3-fold rotation $\chi_{a1}$, $\chi_{a2}$, and $\chi_{a3}$ with three twin domains (Figure S1) ratio $x_1$, $x_2$, and $x_3$ where $x_1+x_2+x_3=1$:

$$\begin{pmatrix} \chi_{a1} \\ \chi_{a2} \\ \chi_{a3} \end{pmatrix} = \chi_{bb} \begin{pmatrix} 0 & 3/4 & 3/4 \\ 3/4 & 0 & 3/4 \\ 3/4 & 3/4 & 0 \end{pmatrix} \begin{pmatrix} x_1 \\ x_2 \\ x_3 \end{pmatrix} \tag{s2}$$

Solving (s2) yields the twin ratio $x_1$, $x_2$, and $x_3$.

## 2. Identification of twined $NiPS_3$ single crystals using polarized Raman measurements

According to the Raman mode assignment reported by Kim *et al.*[1], the low-temperature $P_1$ peak splits into the $A_g$ mode ($P_{1a}$) and the $B_g$ mode ($P_{1b}$), whereas the $P_7$ peak belongs to the $A_g$ symmetry. At room temperature, the splitting between $P_{1a}$ and $P_{1b}$ is smaller than ~1 $cm^{-1}$, which is below our experimental resolution, and therefore the measured $P_1$ peak consists of the unresolved overlap of $P_{1a}$ and $P_{1b}$.

Nevertheless, the room-temperature $P_1$ peak still exhibits a pronounced two-fold-like angular dependence under the parallel linear polarization configuration, indicating that the strong angular anisotropy of the $A_g$ component ($P_{1a}$) remains dominant despite the overlap with the $B_g$ component ($P_{1b}$). By contrast, the $P_7$ peak exhibits much weaker angular dependence and appears nearly isotropic. We therefore attribute the observed two-fold-like angular dependence predominantly to the $A_g$ component ($P_{1a}$), and use this characteristic angular dependence to determine the crystallographic orientation of exfoliated $NiPS_3$ flakes. By correlating Raman results with magnetization measurements, we establish a correspondence between the polarization dependence of the $P_1$ mode and the crystallographic $a$-axis.

Building on this established relationship, we used polarized Raman spectroscopy to verify the presence of twinned domains in our bulk crystals. In some twinned samples, different crystallographic orientations can even be directly identified from the side facets, as shown in Figure S2(c). Such crystals exhibit twinned magnetic behavior, consistent with the magnetization data shown in Figure S2(b). To further verify the presence of twinned domains, polarized Raman measurements were performed on both the top and bottom surfaces of these flakes. As shown in Figure S2(d)-(e), the angular dependence of the $P_1$ mode is rotated by approximately 120° between the two surfaces, consistent with differently oriented twinned domains across the thickness of the flake. Together with the side-facet observations and magnetization measurements, this front–back Raman comparison provides an additional criterion for distinguishing flakes with minimal twinning from twin-rich samples before subsequent physical property measurements.

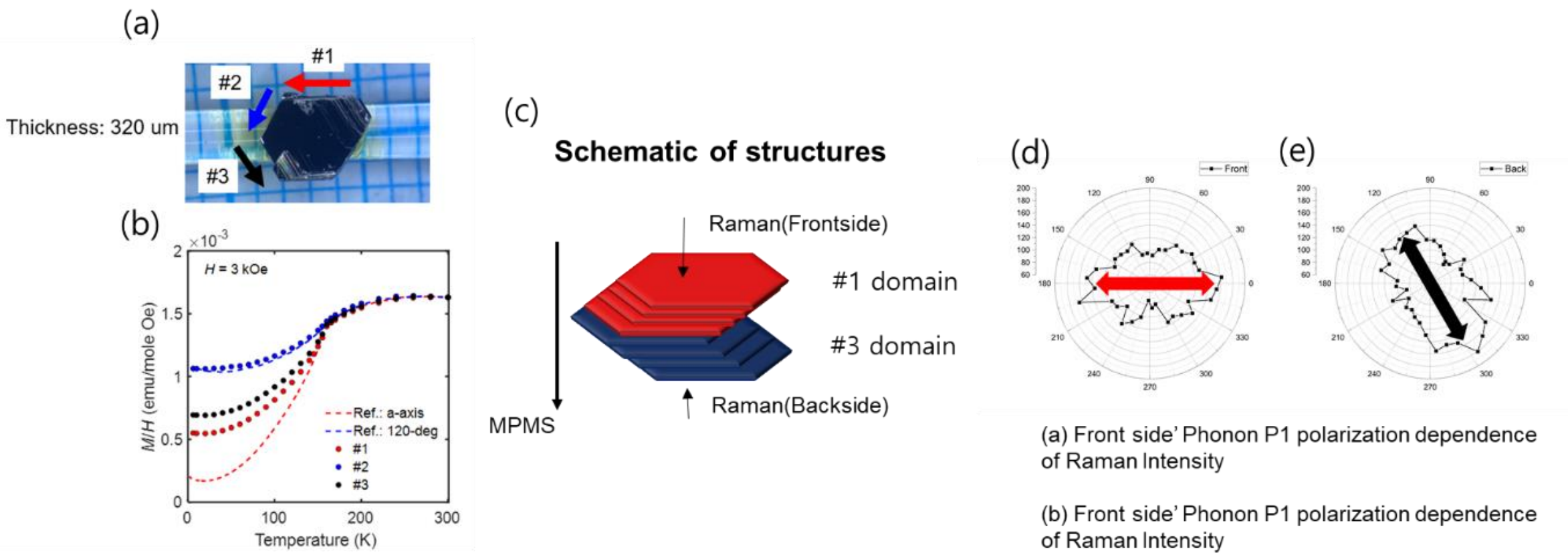

**Figure S2**. Identification of twin-rich $NiPS_3$ flakes by complementary magnetization and polarized Raman measurements. (a) Optical microscope image of a representative twin-rich $NiPS_3$ flake. (b) Angle-dependent magnetization measured with the magnetic field rotated in 120° intervals, showing the characteristic response of a twin-rich sample. (c) Schematic illustration of the side view of the flake, assuming two crystallographic twin domains with different orientations coexist through the thickness. (d,e) Angular dependence of the Raman intensity of the $P_1$ phonon mode measured on the top (d) and bottom (e) surfaces of the same flake using parallel linear polarization with a 532 nm excitation laser. The angular dependence of the $P_1$ mode is rotated by approximately 120° between the two surfaces, consistent with differently oriented twinned domains.

## 3. Comparison between nearly twin-free and twin-rich samples in thermal conductivity

To further investigate the effect of twin domains on the bulk physical properties, we compared the thermal conductivity of the nearly twin-free sample and the twin-rich sample, which exhibit almost identical magnetic susceptibility along three different in-plane directions (Figure S3(a)). As shown in Figure S3(b), we first compared the temperature-dependent thermal conductivity at zero magnetic field. Notably, the peak thermal conductivity of the twin-rich sample is reduced by approximately 30% compared to that of the nearly twin-free sample. This suppression reflects enhanced phonon scattering induced by the higher density of structural domain boundaries in the twin-rich crystal.

Furthermore, the field-dependent thermal conductivity also exhibits clear domain-dependent features (Figure S3 (c)). The twin-rich sample does not exhibit the sharp anomaly near the 10 T spin-flop transition that is clearly visible in the nearly twin-free sample. Instead, a much smaller and broader dip is observed. This broadening can be attributed to macroscopic volume averaging, where the sharp dip originating from the specific domain—where the applied magnetic field is aligned with its crystallographic *a*-axis—is averaged out by signals from other misaligned twin domains within the bulk sample. This macroscopic averaging effect is consistent with the domain sensitivity observed in our magnetization measurements.

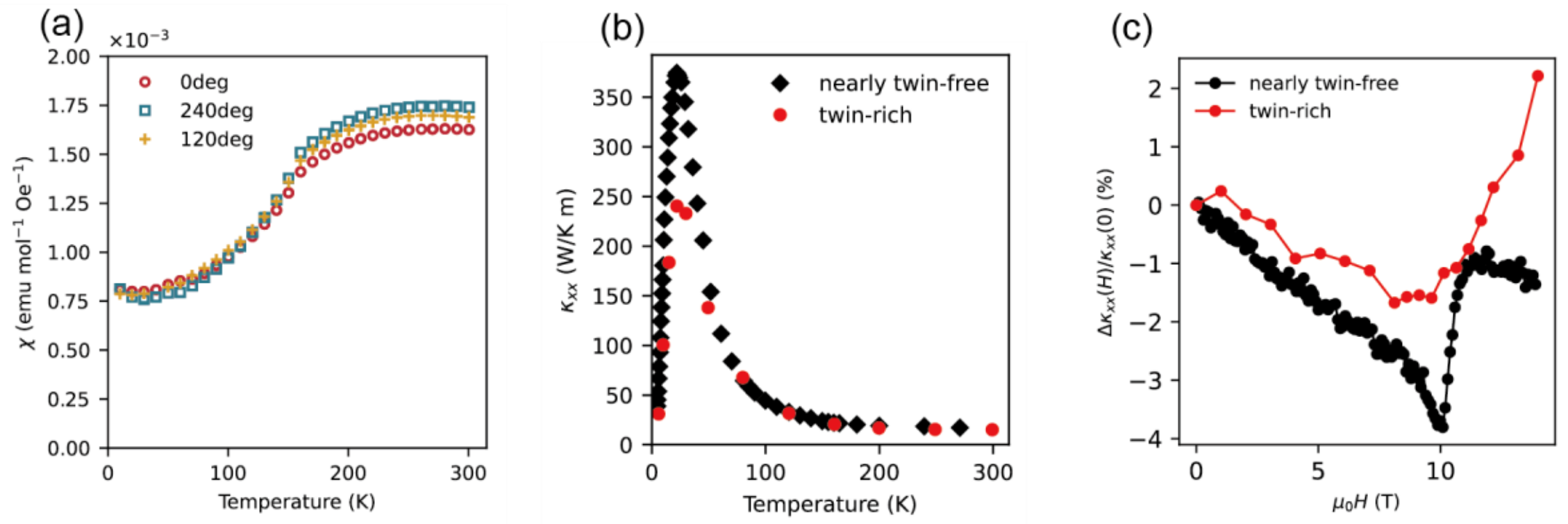

**Figure S3**. (a) Magnetic susceptibility of the twin-rich sample. (b) Temperature-dependent thermal conductivity at zero magnetic field and (c) magnetic field-dependent thermal conductivity at 6 K for the nearly twin-free and twin-rich samples.

### 4. Comparison between nearly twin-free and typically twinned samples in Field-dependent magnetization

To elucidate the influence of crystallographic twinning on the macroscopic magnetic properties, Figure S4 compares the field-dependent magnetization of the nearly twin-free crystal used in the main text with a representative twinned crystal. These experimental results are further compared with theoretical simulations for a perfect single-domain crystal and a representative multidomain configuration.

The field-dependent magnetization is highly sensitive to crystallographic twinning because the spin-flop transition occurs when the magnetic field is applied near the magnetic easy axis of the domain. In a twinned crystal, the measured $M(H)$ curve for a nominal field direction represents a volume-weighted average over multiple monoclinic domains whose crystallographic *a*-axes are rotated by 120° with respect to each other. Consequently, the single-domain response is mixed with contributions from other domains where the applied field is misaligned. This macroscopic averaging broadens the sharp spin-flop anomaly expected for a single-domain crystal.

This trend is consistently observed in both the experimental and simulated data. The nearly twin-free crystal exhibits a sharp spin-flop transition for $H \parallel a$. In contrast, no corresponding transition is observed for $H \parallel b$, which is in excellent agreement with the single-domain calculation as described in the main text. By contrast, the twinned sample displays a more averaged magnetic response with a significantly less sharply resolved spin-flop feature. The corresponding simulation for a representative multidomain mixture (with a 6:2:2 domain population) successfully reproduces this qualitative behavior, confirming that twinning broadens the spin-flop transition and obscures the highly anisotropic intrinsic response by mixing contributions from differently oriented twin domains.

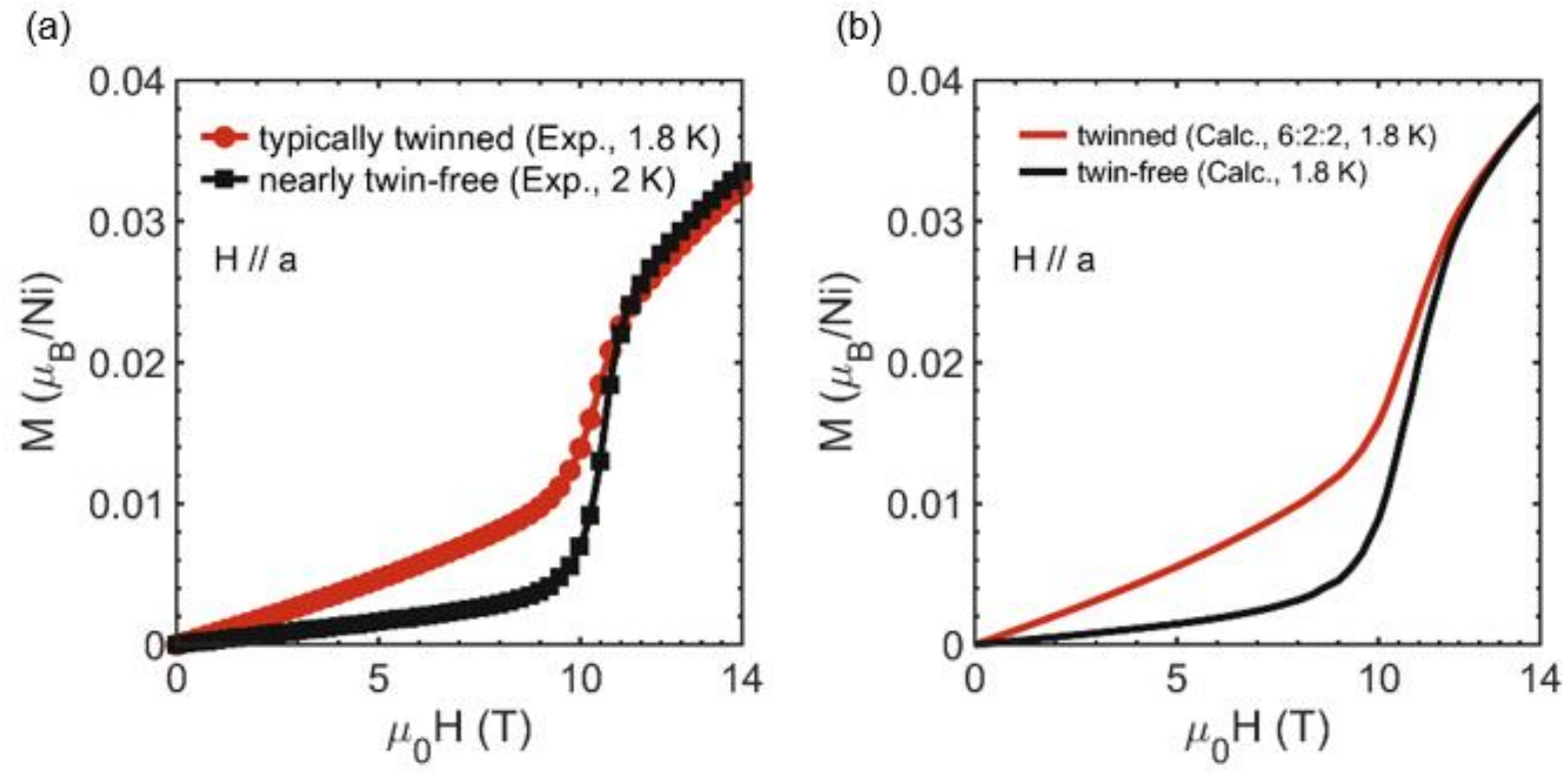

**Figure S4**. (a) Experimental and (b) calculated results of magnetization field dependence for twinned and nearly twin-free crystals.

## 5. Discussion of canting in the magnetic structure

There has been a misrepresentation about the magnetic structure that has gone unnoticed in the literature—the real magnetic structure reported in A. R. Wildes *et al.*[2] has a canting angle of 13.5° out of the ab plane. But a widely used spin model has a moment lying in the *ab* plane, that is, a zero canting angle. For example, two inelastic neutron scattering reports used an uncanted magnetic structure to simplify the analysis and render it analytically accessible. The authors obtained the following single-ion anisotropy parameters using the uncanted model[3]:

$$A_x = -0.010 \pm 0.005 \text{ meV}, \qquad A_z = 0.21 \pm 0.03 \text{ meV}.$$

We think that it is a subtle but important point regarding the magnetic structure. We want to address it by using the real magnetic structure in our calculation. To convert it to the real canted model, we made an axis transformation using a new axis system, the tilda axis, with the $\tilde{x}$ being the correct direction of the canted magnetic moment. A new rotated coordinate system (denoted with tildes) is defined as:

$$\tilde{x} = \cos\theta_A\, \hat{x} - \sin\theta_A\, \hat{z}, \qquad \tilde{z} = \sin\theta_A\, \hat{x} + \cos\theta_A\, \hat{z},$$

with $\theta_A = 13.5°$. Utilizing this properly canted model, the anisotropy parameters were slightly refined to reproduce the observed metamagnetic transition quantitatively. The precise parameters employed in the present work are:

$$A_{\tilde{x}} = -0.009 \text{ meV}, \qquad A_{\tilde{z}} = 0.21 \text{ meV}.$$

Notably, the parameter $A_{\tilde{z}}$ remains identical to the previously reported neutron scattering value, while $A_{\tilde{x}}$ is only marginally adjusted from its central value of -0.01, remaining well within the established experimental uncertainty. This slight refinement in $A_{\tilde{x}}$ provides an improved quantitative reproduction of the experimental spin-flop field observed at 10.5 T for *H* || *a* (Figure S5). Furthermore, the finite pre-spin-flop slope for *H* || *a* is reproduced by the canted-axis model even in a single-domain calculation, indicating that it arises naturally from the canted easy-axis geometry rather than requiring residual twin-domain contributions. We also note that inclusion of spin canting in our simulation model makes a significant difference in the calculated *M*(*H*) curves (Figure S6).

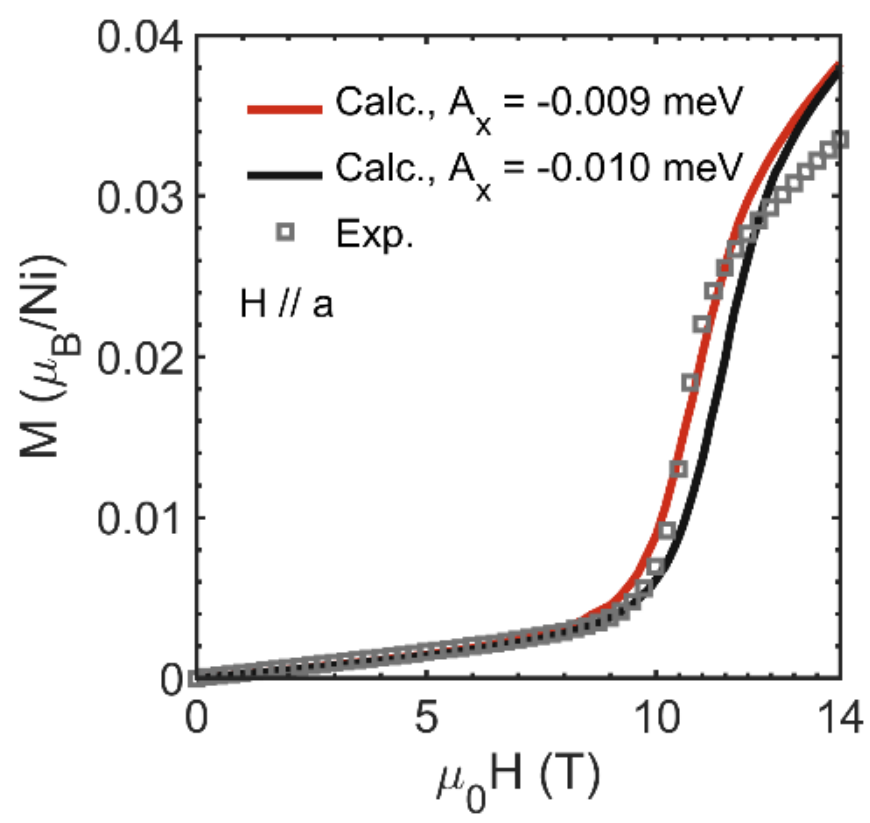

**Figure S5.** *M*(*H*) experimental result and calculations for $A_{\tilde{x}}$=-0.009 meV and -0.01 meV.

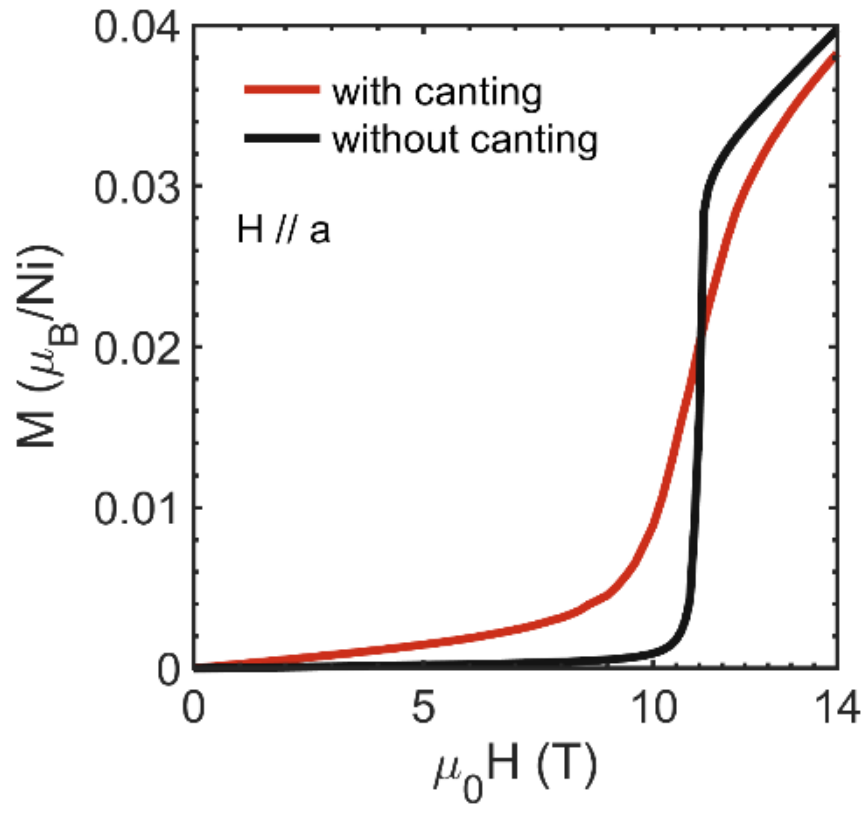

**Figure S6.** Simulated *M*(*H*) curves with and without spin canting in the model.

## 6. Temperature dependence of field-dependent thermal conductivity

It is highly challenging to pinpoint the exact microscopic mechanism underlying the distinct field-dependent slopes of thermal conductivity [4–8]. But these features likely arise from an intricate interplay of primary factors: not only the field-dependent scattering sources of phonon heat transport, but also the role of magnons as active heat carriers, alongside the potential involvement of magnon-phonon hybridization. While the theoretical approach in the main text (Eq. 9) strictly focuses on establishing the simplest possible model capable of capturing the drastic dip driven by the spin-flop transition, here we qualitatively explore the potential mechanisms driving the slope changes at other field regimes by considering a toy model that incorporates magnon heat conduction, which provides an increasing field-dependent contribution. In this model, the total thermal conductivity is expressed as the sum of the phonon and magnon channels:

$$\kappa_{xx}^{total} = \kappa_{xx}^{ph} + \kappa_{xx}^{mag}$$

By introducing a simple magnon thermal conductivity model under the assumption of a constant magnon group velocity and relaxation time, the magnon thermal conductivity is given by:

$$\kappa_{xx}^{mag} \approx C_{mag} v_{mag}^2 \tau_{mag} \sim C_{mag} \approx \sum_{\mathbf{k},mode} E_{mag} \frac{\partial n_{BE}(E_{mag})}{\partial T} \approx \frac{E_{mag}^2}{k_B T^2} n_{BE}(n_{BE}+1)$$

By tuning the parameter $A_{\mathrm{mag}}$ in Eq. 9, the total thermal conductivity can be approximated as:

$$\kappa_{xx}^{total} \approx \kappa_{xx}^{ph}(A_{mag}) + A_1 E_{\mathrm{mag}}^2 n_{\mathrm{BE}}(E_{\mathrm{mag}})(n_{\mathrm{BE}}(E_{\mathrm{mag}}) + 1)$$

Using $E_{\mathrm{mag}}$ as the calculated magnon gap shown in Figure 6(b) of the main text, this model can qualitatively reproduce the temperature evolution of the experimental data. Specifically, it captures the decreasing trend with magnetic field at very low temperatures, which then changes to an increasing trend around 14 K. For the simulations shown in Figure S7, the parameter $A_{\mathrm{mag}}$ was set to $1.0\times10^{-6}$, $2.0\times10^{-5}$, and $1.0\times10^{-4}$ for 6, 10, and 14 K, respectively, while $A_1$ was fixed at 0.02. The large variation in $A_{\mathrm{mag}}$ reflects the substantial temperature-dependent scattering rates that our simplified assumptions do not capture. Developing a more realistic and quantitative model that explicitly incorporates these complex scattering rates is left for future work.

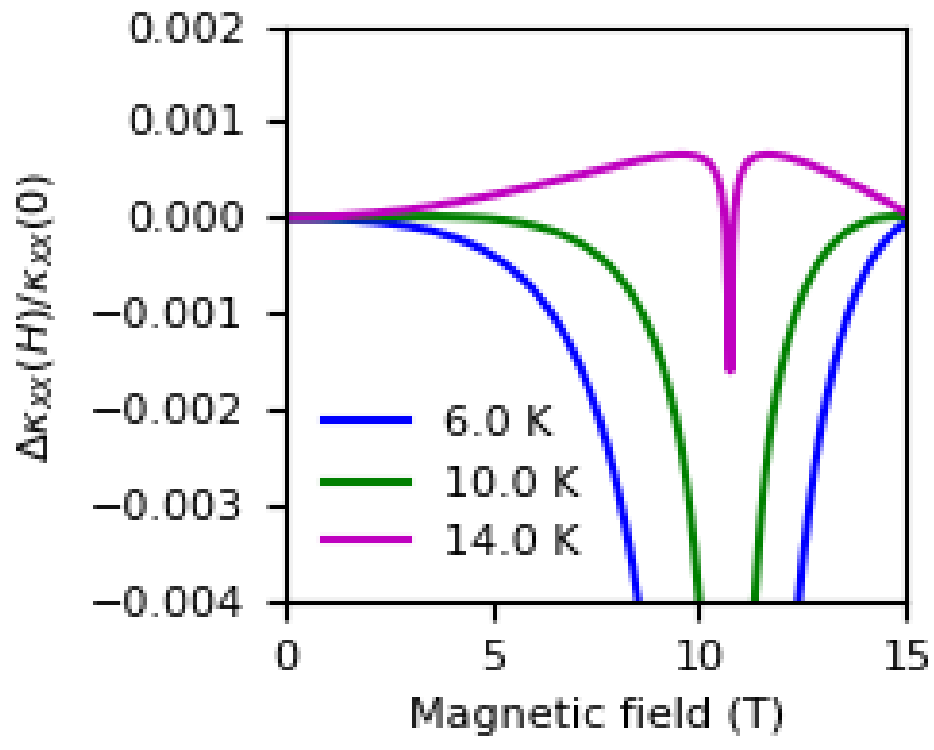

**Figure S7**. Simulated magnetic field dependence of the thermal conductivity at 6, 10, and 14 K based on the proposed toy model.

## 7. Single crystal XRD

Direct crystallographic verification of our mm-scale bulk $NiPS_3$ samples—whose nearly twin-free nature is achievable by magnetization measurements—is severely constrained by the inherent properties of vdW materials. The bulk crystals significantly exceed standard SCXRD beam dimensions, and any mechanical effort to fragment them readily induces artificial deformation and twinning. Furthermore, due to a nearly hexagonal symmetry in the *ab* plane of $NiPS_3$, identifying twins requires an extremely careful approach to resolve highly overlapping peaks.

To obtain qualitative structural evidence for the dominance of a single domain, we conducted comparative SCXRD measurements. We compared a small single crystal, carefully cleaved from the nearly twin-free bulk sample, with an as-grown twinned single crystal. As shown in Figure S8, the reconstructed precession images of the $(h\bar{1}l)$ plane reveal that the selected nearly twin-free crystal exhibits distinct Bragg peaks originating from a single domain (Figure S8(a)). In contrast, the twinned single crystal shows continuous intensity connecting the single domain Bragg peaks along the c*-axis (Figure S8(b)). As supported by our simulation (Figure S8(c)-(d)), these intermediate intensities coincide with the Bragg reflections originating from the other twin domains.

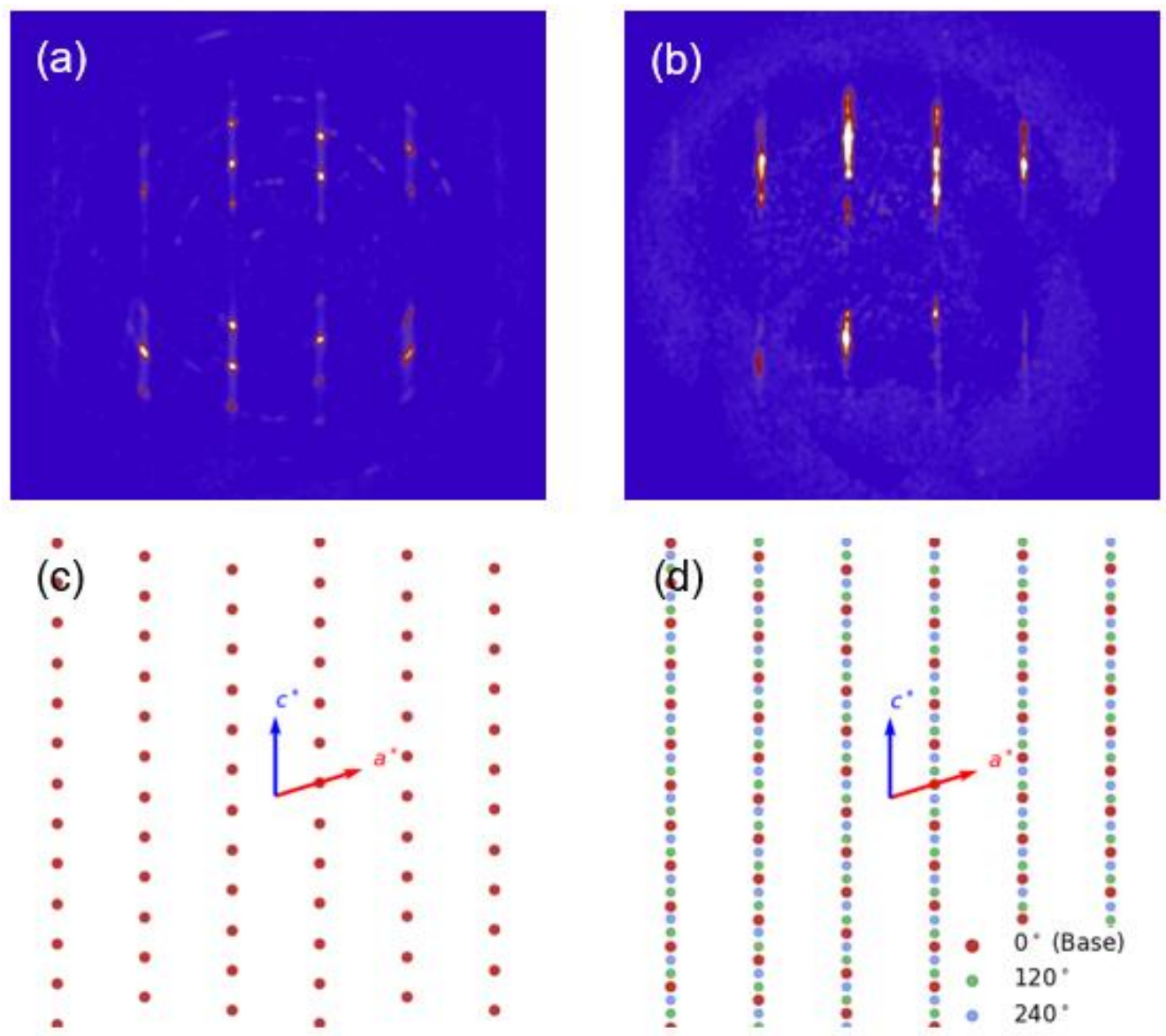

**Figure S8**. Reconstructed precession images of the ($h\bar{1}l$) plane for (a) a $NiPS_3$ single crystal cleaved from a nearly twin-free bulk crystal, and (b) an as-grown twinned $NiPS_3$ single crystal. Simulated Bragg peak positions in the ($h\bar{1}l$) plane for (c) a single domain and (d) three different twin domains.

**References (Supporting Information)**

[1] K. Kim, S. Y. Lim, J.-U. Lee, S. Lee, T. Y. Kim, K. Park, G. S. Jeon, C.-H. Park, J.-G. Park, H. Cheong, *Nat. Commun.* **2019**, *10*, 345.
[2] A. R. Wildes, V. Simonet, E. Ressouche, G. J. McIntyre, M. Avdeev, E. Suard, S. A. J. Kimber, D. Lançon, G. Pepe, B. Moubaraki, T. J. Hicks, *Phys. Rev. B* **2015**, *92*, 224408.
[3] A. Scheie, P. Park, J. W. Villanova, G. E. Granroth, C. L. Sarkis, H. Zhang, M. B. Stone, J.-G. Park, S. Okamoto, T. Berlijn, D. A. Tennant, *Phys. Rev. B* **2023**, *108*, 104402.
[4] D. D. Vu, R. A. Nelson, B. L. Wooten, J. Barker, J. E. Goldberger, J. P. Heremans, *Phys. Rev. B* **2023**, *108*, 144402.
[5] C. A. Pocs, I. A. Leahy, H. Zheng, G. Cao, E.-S. Choi, S.-H. Do, K.-Y. Choi, B. Normand, M. Lee, *Phys. Rev. Res.* **2020**, *2*, 013059.
[6] C. Xu, C. Carnahan, H. Zhang, M. Sretenovic, P. Zhang, D. Xiao, X. Ke, *Phys. Rev. B* **2023**, *107*, L060404.
[7] S. Guang, N. Li, R. L. Luo, Q. Huang, Y. Wang, X. Yue, K. Xia, Q. Li, X. Zhao, G. Chen, H. Zhou, X. Sun, *Phys. Rev. B* **2023**, *107*, 184423.
[8] C. A. Pocs, I. A. Leahy, J. Xing, E. S. Choi, A. S. Sefat, M. Hermele, M. Lee, *Phys. Rev. Res.* **2025**, *7*, L022007.